\documentclass[aps, pra,twocolumn, showpacs,superscriptaddress,10pt,floatfix]{revtex4-1}
\usepackage[english]{babel}
\usepackage[latin1]{inputenc}
\usepackage{hyperref}
\usepackage{amsmath, amssymb, amsfonts}
\usepackage{amssymb,xcolor}
\usepackage{graphicx}
\usepackage{exscale}
\usepackage{latexsym}
\usepackage{cases}
\usepackage{epstopdf}
\usepackage{subfigure}
\usepackage{bm}
\usepackage{color}
\usepackage{amssymb}

\newcommand{\ket}[1]{|#1\rangle}
\newcommand{\bra}[1]{\langle #1|}
\newcommand{\tr}{\mathrm{Tr}}

\newcommand{\n}{\nonumber\\}

\newcommand{\ex}[1]{\langle #1\rangle}

\newcommand{\ii}{ {\rm i} }

\allowdisplaybreaks

\begin{document}

\title{Optimal energy storage in the Tavis-Cummings quantum battery}

\author{Hui-Yu Yang}
\address{School of Physics, Northwest University, Xi'an 710127, China}
\author{Hai-Long Shi}
\email{hl\_shi@yeah.net}
\affiliation{QSTAR and INO-CNR, Largo Enrico Fermi 2, 50125 Firenze, Italy}
\affiliation{Hefei National Laboratory, Hefei 230088, China}
\author{Qing-Kun Wan}
\address{Innovation Academy for Precision Measurement Science and Technology, Chinese Academy of Sciences, Wuhan 430071, China}
\affiliation{University of Chinese Academy of Sciences, Beijing 100049, China}
\author{Kun Zhang}
\email{kunzhang@nwu.edu.cn}
\address{School of Physics, Northwest University, Xi'an 710127, China}
\address{
Shaanxi Key Laboratory for Theoretical Physics Frontiers, Xi'an 710127, China
	}
\address{
Peng Huanwu Center for Fundamental Theory, Xi'an 710127, China
}
\author{Xiao-Hui Wang}
\email{xhwang@nwu.edu.cn}
\address{School of Physics, Northwest University, Xi'an 710127, China}
\address{
Shaanxi Key Laboratory for Theoretical Physics Frontiers, Xi'an 710127, China
	}
\address{
Peng Huanwu Center for Fundamental Theory, Xi'an 710127, China
}
\author{Wen-Li Yang}
\address{
	Institute of Modern Physics, Northwest University, Xi'an 710127, China}
\address{
Shaanxi Key Laboratory for Theoretical Physics Frontiers, Xi'an 710127, China
	}
\address{
Peng Huanwu Center for Fundamental Theory, Xi'an 710127, China
}

\date{\today}

\begin{abstract}
The Tavis-Cummings (TC) model, which serves as a natural physical realization of a quantum battery, comprises $N_b$ atoms as battery cells that collectively interact with a shared photon field, functioning as the charger, initially containing $n_0$ photons.
In this paper, we introduce the invariant subspace method to effectively represent the quantum dynamics of the TC battery.
Our findings indicate that in the limiting case of $n_0\!\gg\! N_b$ or $N_b\!\gg\! n_0$, a distinct SU(2) symmetry emerges in the dynamics, thereby ensuring the realization of optimal energy storage.
We also establish a negative relationship between the battery-charger entanglement and the energy storage capacity.
As a result, we demonstrate that asymptotically optimal energy storage can be achieved in the scenario where $N_b\!=\!n_0\!\gg\! 1$.
Our approach not only enhances our comprehension of the algebraic structure inherent in the TC model but also contributes to the broader theoretical framework of quantum batteries.
Furthermore, it provides crucial insights into the relation between energy transfer and quantum correlations.
\end{abstract}

\maketitle
\section{Introduction}
Embracing the promising trend of device miniaturization, quantum batteries (QBs) have been proposed to exploit quantum features, thereby accelerating charging rates compared to their classical counterparts.
From a theoretical perspective, QBs  offer fundamental insights into the influence of quantum correlations on extractable work, a pivotal concept in quantum thermodynamics\,\cite{PhysRevX.5.041011,PhysRevLett.122.047702,PhysRevLett.129.130602,PhysRevB.104.245418,PhysRevE.87.042123,Allahverdyan_2004,PhysRevE.102.052109,Korzekwa_2016,Caravelli2021energystorage,PhysRevLett.125.180603,PhysRevE.102.042111,PhysRevLett.121.120602,francica2017daemonic,PhysRevE.105.L052101,PhysRevLett.131.030402,PhysRevLett.127.100601}.
QBs, in conjunction with quantum heat engines, offer a tangible approach to incorporate quantum correlations, such as quantum entanglement and quantum coherence, into the field of quantum thermodynamics\,\cite{binder2018thermodynamics,PhysRevLett.89.180402,lostaglio2015description,PhysRevX.5.031044,PhysRevLett.111.250404,Shi_2020,PhysRevA.96.052119,PhysRevLett.128.090602,PhysRevE.90.032102,PhysRevLett.111.230402}.

One notable characteristic of QBs is the phenomenon known as ``charging speedup"\,\cite{PhysRevLett.111.240401,Binder_2015,PhysRevLett.118.150601}.
When connecting a group of $N_b$ QB cells to a shared charger, the charging rate of the QB can potentially scale up to $N_b^2$\,\cite{PhysRevLett.128.140501}, indicating a significant quantum advantage over classical batteries, whose charging speed scales linearly with $N_b$.
This quantum advantage has been verified in various models, including Dicke QBs\,\cite{PhysRevLett.120.117702,PhysRevB.105.115405,zhang2023enhanced}, Sachdev-Ye-Kitaev QBs\,\cite{PhysRevLett.125.236402}, spin-chain QBs\,\cite{PhysRevA.97.022106}, cavity spin-chain QBs\,\cite{PhysRevA.106.032212}, and central-spin QBs\,\cite{PhysRevA.103.052220}.
Nevertheless, it is imperative to emphasize that a charging speedup does not inherently ensure optimal energy storage.
Here, optimal energy storage denotes the complete transfer of input energy from the charger to the battery.
To enrich the theory of QBs, we will shift our focus towards an important cavity quantum electrodynamics (cavity-QED) system, the Tavis-Cummings (TC) model, to address the challenge of achieving optimal energy storage.

\begin{figure*}[htbp]
\includegraphics[width=5.5in]{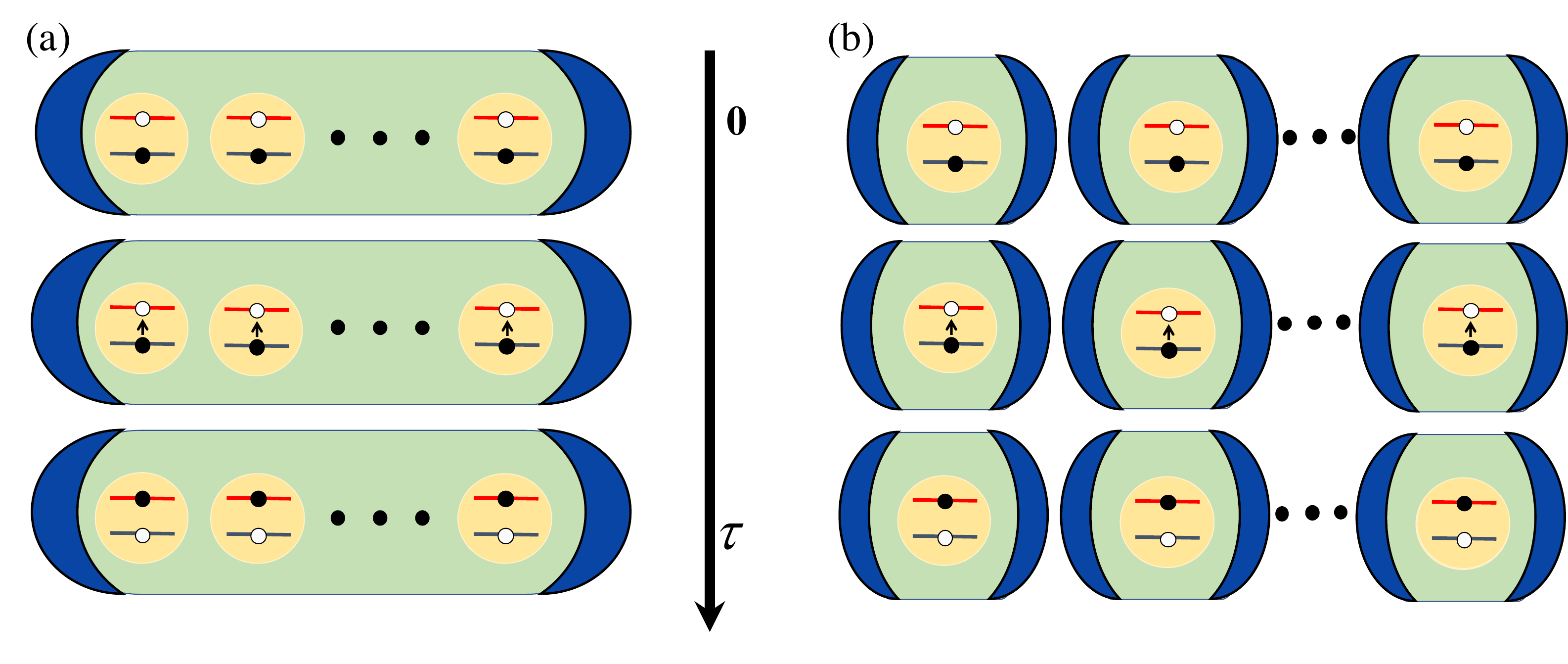}
\caption{
Schematic illustration of two charging schemes for the Tavis-Cummings battery.
(a) Collective charging scheme: $N_b$ battery cells (atoms) commonly interact with a charger (a cavity photon field).
(b) Parallel charging scheme: $N$ single-atom TC models, known as the Jaynes-Cummings model, are placed in parallel.
At $t < 0$, the battery is in the ground state with the lowest energy.
From $0 < t < \tau$, the interaction Hamiltonian $H_I$ is activated to charge the battery.
Finally, at the charging time $\tau$, the interaction is turned off and the energy is stored in the battery.
}\label{fig1}
\end{figure*}

The TC model serves as a foundational framework for investigating the collective behavior of atoms or qubits in the presence of electromagnetic fields\,\cite{PhysRev.170.379,PhysRev.188.692,PhysRevA.8.2517,hepp1973superradiant}.
It constitutes a theoretical underpinning for the fields of quantum optics and cavity QED\,\cite{RevModPhys.93.025005,ZARERAMESHTI20221,garraway2011dicke}.
Furthermore, it enhances our comprehension of qubit manipulation and the generation of entangled states, both of which are pivotal components in the advancement of quantum computing and quantum information processing technologies\,\cite{PhysRevA.68.062316,PhysRevA.75.022312,PhysRevLett.103.083601,PhysRevA.75.022107,steffen2013deterministic,PhysRevLett.124.013601,PhysRevLett.104.100504,higgins2014superabsorption}.
Moreover, the TC model naturally lends itself to describing a QB, wherein each of the $N_b$ atoms serves as an individual battery cell and the optical field serves as the charger, as illustrated in Fig. \,\ref{fig1}(a).
The eigenproblem associated with the TC model can be exactly solved using the algebraic Bethe ansatz\,\cite{bogolyubov2000algebraic}.
However, this approach does not alleviate the inherent challenges, the exponential
wall\,\cite{RevModPhys.71.1253} , associated with calculating quantum dynamics, which is at the core of investigating the energy storage problem.
Transitioning from the eigenproblem to quantum dynamics is inherently challenging, not to mention the complexity of solving the Bethe ansatz equations.
For instance, in Ref.\,\cite{PhysRevA.104.043706} employing the Bethe ansatz approach, numerical calculations were conducted for a scenario involving ten QB cells.
The study concluded that the Fock state is the
optimal initial charger state for the maximum stored energy capacity.
Reference\,\cite{PhysRevLett.122.047702} utilized the {\small{PYTHON}} toolbox {\small{QUTIP}}2 to numerically investigate cases in the TC battery where the initial photon number equals the number of battery cells.
Their findings revealed that a polynomial increase in the support of the battery state guarantees that the stored work can be completely extracted.
For a fixed number of QB cells, Ref.\,\cite{PhysRevLett.129.130602} established a general inverse relationship between the battery-charger entanglement and the extracted work in incoherent QBs.
Despite these productive findings, there remains a need for a deeper understanding of the optimal energy storage problem in the TC battery:
(1) What conditions must be met for a Fock state to achieve optimal energy storage?
(2) How should we address scenarios where the number of battery cells is significantly greater than the initial photon number?
(3) Does the negative relation between the entanglement and the energy storage capacity hold as the number of QB cells varies?

In this paper, we address the aforementioned questions for the TC battery.
In Sec. \ref{TC battery}, we begin by introducing the TC model and proceed to formulate the energy storage problem mathematically.
Subsequently, we introduce the invariant subspace method to overcome the exponential wall obstruction, allowing us to deal with the dynamics of the TC battery when the charger is chosen in a Fock state.
In Sec. \ref{S-Energy}, we first apply the invariant subspace method to analyze the parallel charging scheme.
Subsequently, we delve into three collective charging scenarios: $n_0\!>\!N_b$, $N_b\!>\!n_0$, and $N_b\!=\!n_0$.
Our analysis rigorously establishes that the emergence of SU(2) symmetry in the limit of $n_0\!\gg\! N_b$ or $N_b\!\gg\! n_0$ guarantees the realization of optimal energy storage.
Additionally, we provide a verification of the negative correlation between the entanglement and the energy storage capacity.
Furthermore, we present numerical evidence that, in the limit of $N_b\!=\!n_0\!\gg\! 1$,  asymptotically optimal energy storage can be attained.
In Sec. \ref{Dis}, we provide a discussion regarding the experimental implementation of optimal energy storage within the TC battery and make a conclusion.

\section{Tavis-Cummings battery}\label{TC battery}
The TC battery consists of $N_b$ battery cells, represented by the total spin operators   $S^{\alpha}\!=\!\sum_{j=1}^{N_b}\sigma_j^\alpha/2$ $(\alpha\!=\!x,y,z)$, and a charger described by  bosonic operators $a$ and $\hat a$.
The TC Hamiltonian is given by
\begin{eqnarray}\label{TCB}
&&H=H_b+H_c+\lambda(t) H_I,\\
&&H_b=\omega_bS^{z},\quad H_c=\omega_{c}a^\dag a,\n
&&H_I=g(S^{+}a+S^{-}a^\dag),\nonumber
\end{eqnarray}
where $H_b$, $H_c$, and $H_I$ correspond to the Hamiltonian of the battery, charger and their interaction, respectively.
Here, the parameter $g$ characterizes the flip-flop interaction and $S^\pm\!=\!S^x\pm iS^y$ represents the spin ladder operators.
We always consider the resonance condition, specifically $\omega_b\!=\!\omega_c\!\equiv\! \omega$, thereby guaranteeing the conservation of the total energy, i.e., $[H,H_b+H_c]\!=\!0$.

The QB is initially prepared in a product state $\ket{\psi(0)}\!=\!\ket{0}_b\ket{n_0}_c$, where the battery is in the ground state of $H_b$, i.e., $\ket{0}_b\equiv\ket{\downarrow,\ldots,\downarrow}$ and the charger is in a Fock state $\ket{n_0}_c$.
At time $t>0$, the interaction term $H_I$ is activated to facilitate the transfer of energy from the charger to the battery.
Over a time interval $[0,\tau]$, our objective is to maximize the amount of energy stored in the battery.
This optimal time duration $\tau$ is referred to as the ``charging time."

A pivotal step in our analysis involves the calculation of quantum dynamics for the battery state.
Due to the U(1) symmetry $[H,S^z\!+\!a^\dag a]\!=\!0$ and the local spin-sector symmetry $[H,\vec S^2]\!=\!0$, we observe that the TC battery dynamics can be effectively reformulated within the following subspace,
\begin{eqnarray}\label{Invariant-subspace}
\mathcal B\!=\!{\rm span}\{\ket{0}_b\ket{n_0}_c,
\ket{1}_b\ket{n_0\!-\!1}_c,\cdots,
\ket{d}_b\ket{n_0\!-\!d}_c\},
\end{eqnarray}
where $\ket{m}_b$ denotes the Dicke state $\ket{N_b/2,-N_b/2+m}_b$, with $m=0,\ldots,N_b$ for the battery and $\ket{n}_c$ signifies the Fock state for the charger.
The parameter $d$ is defined as $d\!=\!\min\{N_b,n_0\}$.

Within the basis (\ref{Invariant-subspace}), we can express the Hamiltonian (\ref{TCB}) as a $(d+1)\times(d+1)$ matrix (details are available in Appendix A),
\begin{eqnarray}\label{Effective-H}
\bm H=\begin{pmatrix}
0&u_1&\\
u_1&0&u_2&\\
&\ddots&\ddots&\ddots&\\
&&u_{d-1}&0&u_{d}\\
&&&u_{d}&0
\end{pmatrix},
\end{eqnarray}
where $u_j\!=\!g\sqrt{j(N_b-j+1)(n_0-j+1)}$.
Assuming that $\bm H$ is diagonalized by a unitary matrix $\bm U$, i.e., $\bm H=\bm U \bm E\bm U^\dag$ where $\bm E$ is a diagonal matrix, the matrix representation of the wave function for the entire system at time $t$ is given by
\begin{eqnarray}\label{WaveFunc}
\bm{\psi}(t)=\bm{U} e^{-i\bm Et}\bm U^\dag (1\ 0\ \ldots\ 0)^T.
\end{eqnarray}
Consequently, the corresponding quantum state is
\begin{eqnarray}\label{total-state}
\ket{\psi(t)}=\bm\psi_1(t)\ket{0}_b\ket{n_0}_c+\cdots+\bm\psi_{d+1}(t)\ket{d}_b\ket{n_0-d}_c.\n
\end{eqnarray}
This leads to the reduced density matrix of the battery state:
\begin{eqnarray}\label{RDM}
\rho_b(t)&\equiv& \tr_c(\ket{\psi(t)}\bra{\psi(t)})\n
&=&|\bm \psi_1(t)|^2\ket{0}\bra{0}+\cdots+|\bm\psi_{d+1}|^2\ket{d}\bra{d}.
\end{eqnarray}
At time $t$, the energy stored in battery is given by
\begin{eqnarray}\label{Energy}
\Delta E(t)&\equiv&\tr(H_b\rho_{b}(t))-\tr(H_{b}\rho_{b}(0))\n
&=&\frac{N_b}{2}\omega+\sum_{j=1}^{d+1} \left(j-\frac{N_b}{2}-1\right)|\bm \psi_j(t)|^2\omega.
\end{eqnarray}

The definition of \emph{optimal energy storage} encompasses the following two scenarios:
\\(I) The battery absorbs all the energy initially stored in the charger, resulting in zero charger energy after the charging process.
\\(II) The battery absorbs some energy from the charger and reaches a fully charged state $\ket{\uparrow ,\ldots,\uparrow}_b$.
\\
Case I corresponds to a situation where the initial energy of the charger is less than the energy storage capacity of the battery, i.e., $n_0\!<\! N_b$, while case II corresponds to the opposite case where $n_0\!>\! N_b$.
Referring to Eq. \,(\ref{total-state}), optimal energy storage implies that the evolved quantum state is $\ket{\psi(\tau)}=\ket{d}_b\ket{n_0-d}_c$.
In other words, the optimal condition can also be expressed as
\begin{eqnarray}
\exists \tau\ {\rm such\ that}\ \Delta E(\tau)=\omega d,
\end{eqnarray}
where $d\!=\!\min\{N_b,n_0\}$ and $\Delta E(\tau)\equiv\max_t\Delta E(t)$ is called the energy storage capacity.
From the perspective of quantum correlations, thus, optimal energy transport occurs when there is no entanglement between the battery and the charger.
We will further elaborate on this point in the ensuing discussion.
It is worth noting that the dimension of the matrix \,(\ref{Effective-H}) no longer exhibits exponential growth with the number of battery cells $N_b$ but rather depends linearly on the minimum of $N_b$ and $n_0$.
As a result, compared to the Bethe ansatz approach\,\cite{PhysRevA.104.043706}, our method offers greater numerical tractability.
In the subsequent sections, we will demonstrate that our approach also provides profound insights into the algebraic structure of the TC model, enabling us to analytically address the optimal energy storage problem.

\section{energy transport and Entanglement}\label{S-Energy}
\subsection{Parallel charging: \texorpdfstring{$N_b=1$}{Nb=1}}
The parallel charging scheme involves the assembly of $N$ Jaynes-Cummings (JC) models, each of which contains only a single battery cell, as depicted in Fig. \,\ref{fig1}(b).
The JC model is a special case of the TC model by taking $N_b\!=\!1$.
In this scenario, a straightforward calculation yields the following evolved quantum state \,(\ref{total-state}):
\begin{eqnarray}
\ket{\psi(t)}\!=\!\cos(g\sqrt{n_0}t)\ket{0}_b\ket{n_0}_c\!-\!\ii\sin(g\sqrt{n_0}t)\ket{1}_b\ket{n_0\!-\!1}_c.\nonumber
\end{eqnarray}
It thus follows from Eq.\,(\ref{Energy}) that the energy stored in the battery is given by
\begin{eqnarray}
\Delta E(t)=\omega\sin^2(g\sqrt{n_0}t),
\end{eqnarray}
which indicates that the battery state can be excited to its highest-energy state, $\ket{\uparrow}$, at time $\tau=\pi/(2g\sqrt{n_0})$.
Hence, we can conclude that the optimal energy storage can be achieved in the parallel charging scheme.
However, we know that the collective charging scheme offers a charging speedup\,\cite{PhysRevLett.111.240401,Binder_2015,PhysRevLett.118.150601}.
Therefore, our primary focus will be on seeking optimal energy storage within the collective charging scheme, i.e., when $N_b\!>\!1$.

\subsection{Collective charging: \texorpdfstring{$n_0\!>\!N_b$}{Nb<=n0}}\label{B}
\begin{figure}[t]
\includegraphics[width=3.2in]{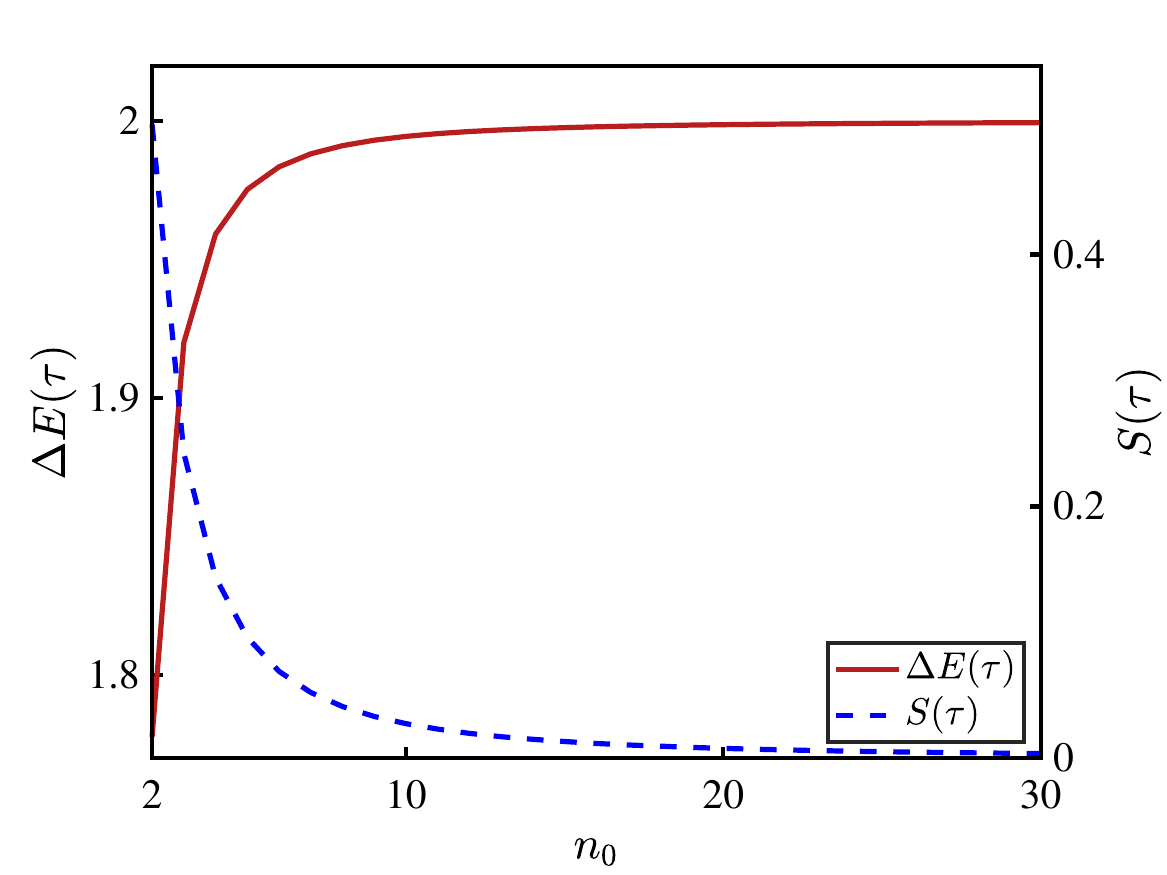}
\caption{
Negative relation between the battery-charger entanglement $S(\tau)$ and the energy storage capacity $\Delta E(\tau)$ in the case of $n_0\!\geq\!N_b\!=\!2$.
Other parameters are set to  $\omega\!=\!1$ and $g\!=\!1$.
}\label{fig2}
\end{figure}

In this section, we consider the case where $n_0>N_b$, with the expectation that the battery can be fully charged.
For this purpose, we first examine the simplest scenario of $N_b=2$ and $n_0> N_b$ to illustrate our analytical approach.
Under this circumstance, the Hamiltonian (\ref{Effective-H}) can be represented as a three-dimensional matrix:
\begin{eqnarray}
\bm H=\begin{pmatrix}
0&g\sqrt{2n_0}&0\\
g\sqrt{2n_0}&0&g\sqrt{2(n_0-1)}\\
0&g\sqrt{2(n_0-1)}&0\\
\end{pmatrix}.
\end{eqnarray}
Its eigenproblem \,(\ref{WaveFunc}) can be solved straightforward as
$\bm{E}={\rm diag}(0,\delta,-\delta)$ and
\begin{eqnarray}
&&\bm U\!=\!\frac{1}{2\sqrt{2n_0-1}}\begin{pmatrix}
2\sqrt{n_0-1}& \sqrt{2n_0}&\sqrt{2n_0}\\
0&\sqrt{4n_0-2}&-\sqrt{4n_0-2}\\
-2\sqrt{n_0}&\sqrt{2(n_0-1)}&\sqrt{2(n_0-1)}
\end{pmatrix},\n
\end{eqnarray}
where $\delta=g\sqrt{4n_0-2}$.
Substituting the above solution into Eq. \,(\ref{WaveFunc}), we obtain
\begin{eqnarray}\label{rho-elements}
|\bm \psi_1(t)|^2&=&\frac{1}{(2n_0-1)^2}[n_0-1+n_0\cos(\delta t)]^2,\n
|\bm \psi_2(t)|^2&=&\frac{n_0}{2n_0-1}[1-\cos^2(\delta t)],\n
|\bm \psi_3(t)|^2&=&\frac{n_0(n_0-1)}{(2n_0-1)^2}[1-\cos(\delta t)]^2.
\end{eqnarray}
Then, according to Eq. \,(\ref{Energy}), we have
\begin{eqnarray}\label{DeltaE_Nb_2}
\Delta E(t)&=&\frac{\omega}{(2n_0-1)^2}[-n_0\cos^2(\delta t)\n
& &+4n_0(1-n_0)\cos(\delta t)+n_0(4n_0-3)].
\end{eqnarray}
The maximum of $\Delta E(t)$ is thus given by
\begin{eqnarray}\label{E2}
\Delta E(\tau)=\omega\left[2-\frac{2}{(2n_0-1)^{2}}\right],
\end{eqnarray}
occurring at $\tau=\pi/[g\sqrt{4n_0-2}]$.
From Eq. \,(\ref{E2}), it is evident that the energy storage capacity $\Delta E(\tau)$ becomes larger as the number of initial photons, $n_0$, increases.
If $n_0\to\infty$, optimal energy storage can be realized.

As demonstrated in Ref. \,\cite{PhysRevLett.129.130602}, the energy storage is constrained by the battery-charger entanglement.
Hence, it is anticipated that as $n_0$ increases, the entanglement will decrease and eventually vanish as $n_0\to\infty$.
Utilizing the von Neumann entropy and the expression for the reduced density matrix of the battery in Eq. \,(\ref{RDM}), we find that at time $\tau$, the entanglement entropy is given by
\begin{eqnarray}
S(\tau)&\equiv&-\tr(\rho_b(\tau)\log_2\rho_b(\tau))\n
&=&h\left(\frac{1}{(2n_0-1)^{2}}\right)
\end{eqnarray}
where $h(x) =-x\log_2x -(1 - x)\log_2(1 - x)$ represents the binary Shannon entropy function.
Since $h(x)$ is an increasing function when $0\leq x\leq 1/2$ and $n_0> N_b=2$, we can conclude that entanglement decreases as $n_0$ increases.
Therefore, we validate the negative relation between the energy storage capacity and the entanglement (see Fig. \,\ref{fig2}).

\begin{figure}[t]
\includegraphics[width=3.2in]{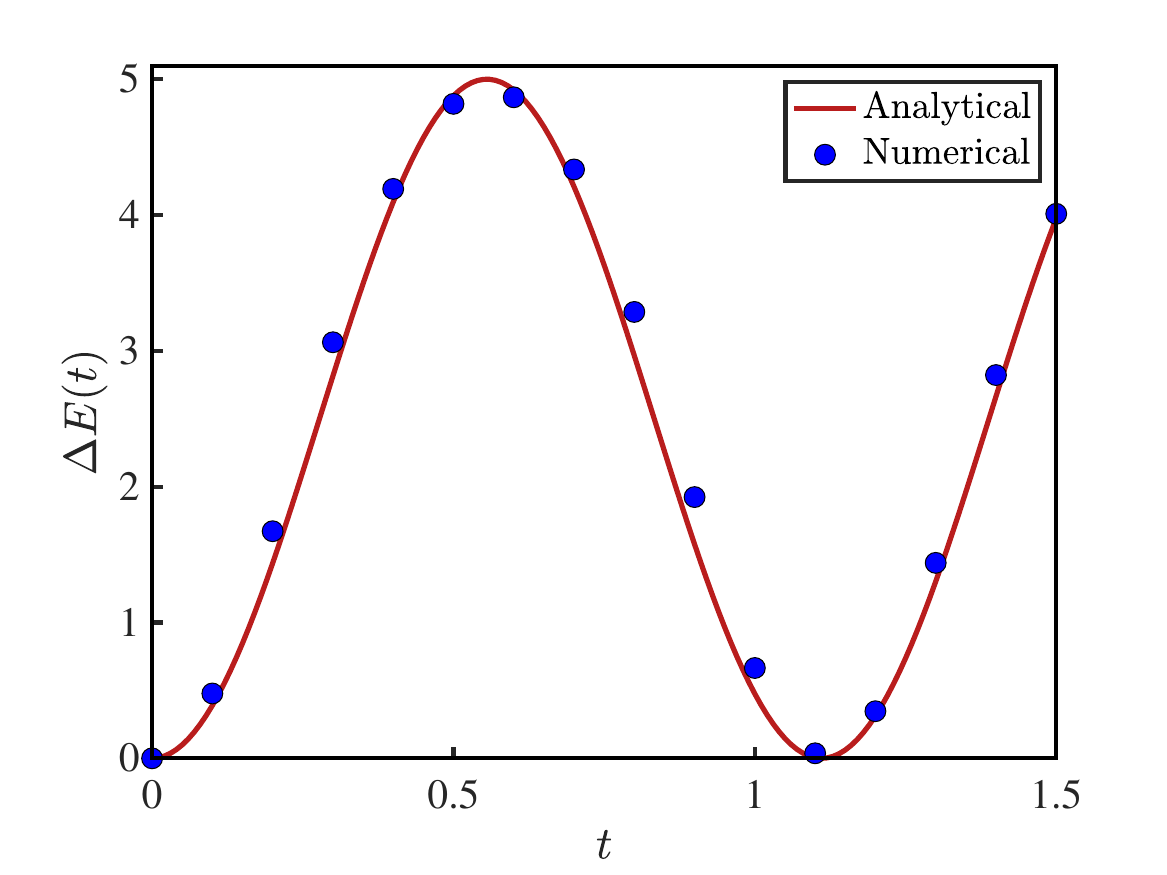}
\caption{
Comparison of analytical results of Eqs. \,(\ref{Delta-E-1}) and (\ref{Delta-E-2}) with the numerical results.
This comparison is valid for both the cases of $n_0\!=\!N_b/2\!=\!5$ and $N_b\!=\!n_0/2\!=\!5$.
Other parameters are set to $\omega\!=\!1$ and $g\!=\!1$.
}\label{fig3}
\end{figure}

A natural question that arises is whether optimal energy storage can be achieved under the condition $n_0\!\gg\! N_b$, not limited to the specific case of $N_b\!=\!2$ as discussed earlier.
It appears challenging to address this question by directly solving the eigenproblem of Eq. \,(\ref{Effective-H}) for arbitrary values of $N_b$.
Fortunately, we observe that under the condition $n_0\!\gg\! N_b$ the matrix element $u_j\!=\!g\sqrt{j(N_b-j+1)(n_0-j+1)}$ in Eq. \,(\ref{Effective-H}) can be approximated as $u_j\!\simeq\! g\sqrt{n_0-N_b/2+1/2}\sqrt{j(N_b-j+1)}$.
With this approximation, the Hamiltonian \,(\ref{Effective-H}) can be mapped to the spin-$N_b/2$ operator (see Appendix B),
\begin{eqnarray}\label{H-limit-1}
\bm{H}\simeq \Omega\bm J^x,
\end{eqnarray}
where $\Omega$ is the generalized Rabi frequency:
\begin{eqnarray}\label{Omega-1}
\Omega=2g\sqrt{n_0-\frac{N_b-1}{2}}.
\end{eqnarray}
Now, the Hamiltonian Eq. \,(\ref{H-limit-1}) is just one of the generators of the su(2) Lie algebra. Therefore, we can assert that SU(2) symmetry emerges in this limiting case.
By using Eq. \,(\ref{H-limit-1}), we can reformulate Eq. \,(\ref{Energy}) as
\begin{eqnarray}\label{Energy-2}
\Delta E&=&\frac{N_b\omega}{2}+\bm{\psi}^\dag(0)\bm{J}^z(t)\bm{\psi}(0)\omega,
\end{eqnarray}
where $\bm{J}^z(t)=\exp(\ii t\Omega \bm{J}^x)\bm{J}^z\exp(-\ii t\Omega \bm{J}^x)$.

Now, it is evident that our question has been reduced to a Lie algebra problem, owing to the emergence of SU(2) symmetry in the limit $n_0\!\gg\! N_b$.
Using the Baker-Campbell-Hausdorff formula, we have
\begin{eqnarray}\label{BCH}
\bm J^z(t)
=\sin(\Omega t)\bm J^y+\cos(\Omega t)\bm J^z.
\end{eqnarray}
By substituting Eq. \,(\ref{BCH}) into Eq. \,(\ref{Energy-2}), we obtain
\begin{eqnarray}\label{Delta-E-1}
\Delta E(t)=\frac{N_b\omega}{2}[1-\cos(\Omega t)],
\end{eqnarray}
which is extremely accurate even in the finite case $N_b\!=\!n_0/2\!=\!5$ (see Fig. \,\ref{fig3}).
Equation (\ref{Delta-E-1}) indicates that optimal energy storage $\Delta E=N_b\omega$ can be achieved at $\tau=\pi/\Omega$ when $n_0\!\gg\! N_b$.

\subsection{Collective charging: \texorpdfstring{$N_b>n_0$}{Nb>n0}}

In Sec. \,\ref{B}, we employed the invariant subspace method, as given in Eq. \,(\ref{Effective-H}), to reveal the fundamental algebraic structure that characterizes the quantum dynamical behavior of the TC model.
This approach was entirely applicable to the case where $N_b>n_0$.
To delve deeper into the core of our investigation, we will now proceed directly to explore the scenario where $N_b\gg n_0$ and claim that optimal energy storage can be realized.

For $ N_b\!\gg\! n_0$, we can approximate $u_j$ as $u_j\!\simeq\! g\sqrt{N_b-n_0/2+1/2}\sqrt{j(n_0-j+1)}$ and then the Hamiltonian \,(\ref{Effective-H}) can be approximated as a spin-$n_0/2$ operator $\bm {\tilde J}^\alpha$, i.e.,
\begin{eqnarray}\label{H-limit-2}
\bm{H}\simeq \tilde \Omega\bm {\tilde J}^x,
\end{eqnarray}
where
\begin{eqnarray}\label{Omega-2}
&&\tilde \Omega=2g\sqrt{N_b-\frac{n_0-1}{2}}.
\end{eqnarray}
Thus, the energy transfer Eq. \,(\ref{Energy}) can be repressed as
\begin{eqnarray}\label{Energy-3}
\Delta E&=&\frac{n_0\omega}{2}+\bm{\psi}^\dag(0)\bm{\tilde J}^z(t)\bm{\psi}(0)\omega,
\end{eqnarray}
where $\bm{\tilde J}^z(t)=\exp(\ii t\tilde \Omega \bm{\tilde J}^x)\bm{\tilde J}^z\exp(-\ii t\tilde \Omega\bm{\tilde J}^x)$.
We see that these equations are highly similar to the $n_0\gg N_b$ case if we exchange $n_0$ and $N_b$.
Therefore, we can deduce that
\begin{eqnarray}\label{Delta-E-2}
\Delta E(t)=\frac{n_0\omega}{2}[1-\cos(\tilde \Omega t)],
\end{eqnarray}
which suggests that optimal energy storage $\Delta E=n_0\omega$ can be realized at charging time $\tau=\pi/\tilde \Omega$ in the case of $N_b\gg n_0$.
Figure \,\ref{fig3} shows the consistency of the limit result with the numerical result obtained also in the finite case $n_0\!=\!N_b/2\!=\!5$.

\subsection{Collective charging: \texorpdfstring{$n_0=N_b$}{Nb<=n0}}
The most economical charging scheme is preparing the initial charger with the same energy as the energy storage capacity of the battery, i.e., $n_0\!=\!N_b$.
As we can see in the case of $n_0\!\geq\!N_b\!=\!2$ (Fig.
\,\ref{fig2}), however, the energy storage capacity is weakest for the $n_0\!=\!N_b\!=\!2$ case.
Hence, we inquire whether, in the limiting case of $n_0=N_b\gg 1$, we can approach optimal energy storage.
Within this parameter range, no evident approximations for Eq. \,(\ref{Effective-H}) can be employed to obtain analytical results.
Nevertheless, we can also perform an effective numerical calculation based on Eq. \,(\ref{Effective-H}) to analyze the charging behavior in the case of $n_0\!=\!N_b$.

\begin{figure}[t]
\includegraphics[width=3.4in]{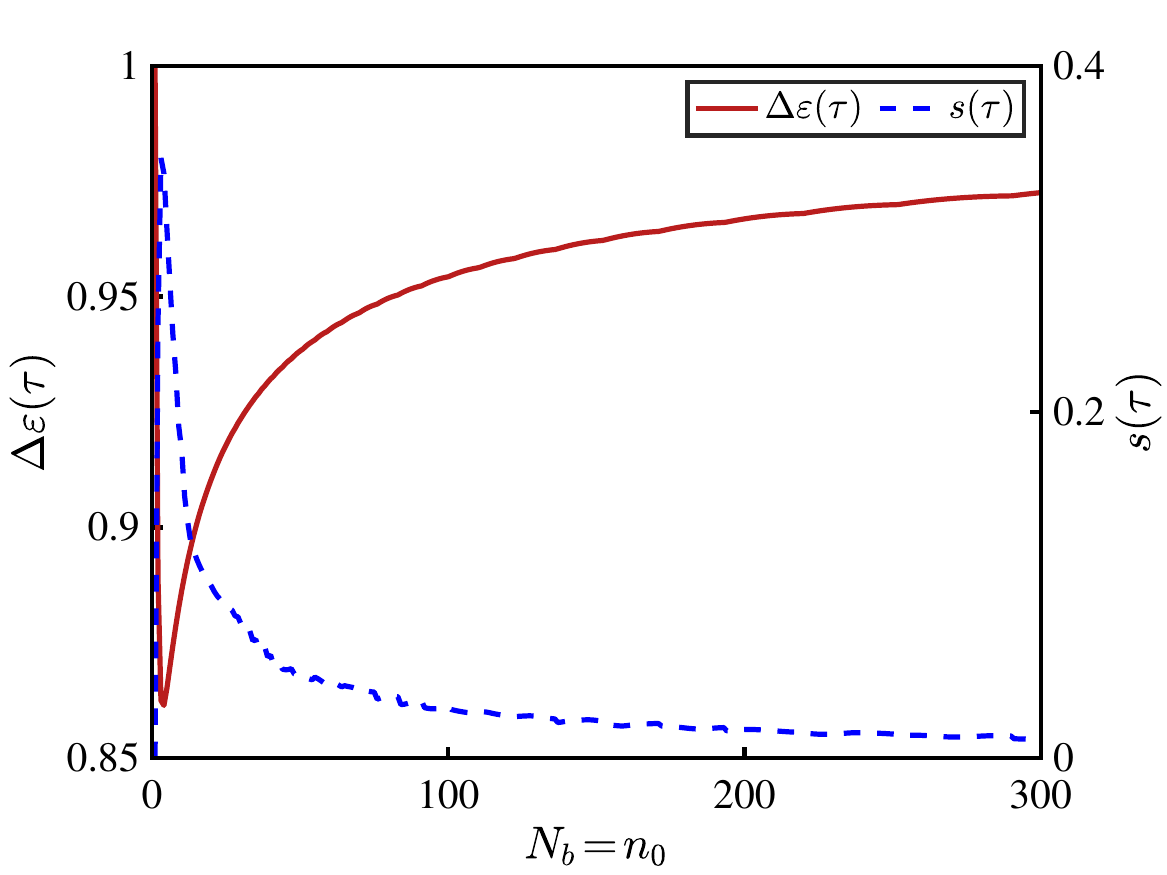}
\caption{
Negative relation between average entanglement $s(\tau)\equiv S(\tau)/N_b$ and average energy storage capacity $\Delta \varepsilon(\tau)\equiv\Delta E(\tau)/N_b$ in the case of $N_b\!=\!n_0$.
Other parameters are set to $\omega\!=\!1$ and $g\!=\!1$.
}\label{fig4}
\end{figure}

Figure \,\ref{fig4} illustrates the average entanglement $s(\tau)\!\equiv\! S(\tau)/N_b$ and the average stored energy $\Delta \varepsilon(\tau)\!\equiv\!\Delta E(\tau)/N_b$ at the charging time $\tau$ for the scenario where $N_b\!=\!n_0$.
The motivation for averaging these quantities stems from the fact that the support of the reduced density matrix, Eq. \,(\ref{RDM}), is determined by $\min\{n_0,N_b\}+1$, which equals $(N_b+1)$ in this scenario.
Therefore, average quantities enable us to compare cases with varying values of $N_b$.
Figure \,\ref{fig4} shows that the average energy storage capacity $\Delta \varepsilon(\tau)$ initially decreases and then increases with an increase in the size of the QB.
Conversely, the average entanglement between the battery and the charger displays opposite behavior compared to
$\Delta \varepsilon(\tau)$.
We have previously established a negative relation between the energy storage and the entanglement for a fixed QB size in Ref. \,\cite{PhysRevLett.129.130602}.
Figure \,\ref{fig4} demonstrates that this negative relation persists even when the QB size is not fixed.
%
It is important to emphasize that as $N_b$ becomes sufficiently large $(>100)$, the asymptotically optimal energy transfer $[\Delta \varepsilon(\tau)>0.95]$ can be achieved, as depicted in Fig. \,\ref{fig4}.
It is worth noting that a specific perturbation theory, based on polynomially deformed su(2) algebras\,\cite{PhysRevA.67.053808}, can be applied to address the eigenproblem of the TC model in the case of $N_b=n_0$.
Exploring the potential applicability of this method to address QB problems is an interesting topic for further discussion.

\section{Discussion}\label{Dis}
In this paper, we have investigated the realization of optimal energy storage in the TC battery.
By selecting the Fock state as the initial state of the charger, we have introduced the invariant subspace method to investigate the quantum dynamics of the TC model.
This approach has enabled us to demonstrate that SU(2) symmetry emerges in cases where either $n_0\gg N_b$ or $N_b\gg n_0$.
Consequently, under these two conditions, optimal energy storage in the TC battery can be achieved.
In the scenario where $N_b\!=\!n_0$, we have conducted numerical investigations into the energy storage problem, confirming the existence of a negative relation between the battery-charger entanglement and the energy storage capacity.
Our numerical results have indicated the possibility of asymptotically achieving optimal energy storage in the limit $N_b\!=\!n_0\!\gg\! 1$.

It is worth noting that the TC battery is possibly realized using the state-of-the-art solid-state technology through the circuit/cavity QED\,\cite{PhysRevLett.103.083601,PhysRevApplied.14.024025,PhysRevB.79.180511,PhysRevLett.114.183601,PhysRevLett.130.173601}.
Reference\,\cite{PhysRevLett.103.083601} reported an ideal realization of the TC model in the absence of atom number and
coupling fluctuations by embedding a discrete number of fully controllable superconducting qubits at
fixed positions into a transmission line resonator.
Reference\,\cite{PhysRevLett.130.173601} realized strong coupling between deterministic single-atom arrays and a high-finesse miniature optical cavity.
While our theoretical starting point has been the Fock state as the initial charge, in experimental settings, coherent light can be used as an approximation to the Fock state.
This simplifies the difficulty of experimentally preparing Fock states.
Consequently, our theoretical approaches to optimal energy storage can be implemented in a well-designed circuit/cavity QED system.
Our work not only contributes to a profound understanding of the algebraic structure of the TC model but also offers significant insights into the problem of optimal energy storage.
We anticipate that our approach can be extended to address other quantum battery-related challenges in the future.

\begin{acknowledgments}
This work was supported by the NSFC (Grants No. 12275215, No. 12305028, No. 12247103, and No. 12175178), the Major Basic Research Program of Natural Science of Shaanxi Province (Grant No. 2021JCW-19), Shaanxi Fundamental Science Research Project for Mathematics and Physics (Grant No. 22JSZ005) and the Youth Innovation Team of Shaanxi Universities.
Q.-K.W. and H.-L.S. received partial support from the NSFC Key Grants No. 12134015 and No. 92365202.
H.-L.S. was supported by the European Commission through the H2020 QuantERA ERA-NET Cofund in Quantum Technologies project ``MENTA.''

H.-L.S. and H.-Y.Y. contributed equally to this work.

\end{acknowledgments}

\section*{Appendix A: Matrix expression of Hamiltonian (\ref{TCB}) in the invariant subspace}
According to the algebraic relations,
\begin{eqnarray}
&&S^+\ket{m}_b\!=\!\sqrt{(N_b-m)(m+1)}\ket{m+1}_b,\n
&&S^-\ket{m}_b\!=\!\sqrt{m(N_b-m+1)}\ket{m-1}_b,\n
&&a^\dag\ket{n}_c=\sqrt{n+1}\ket{n+1}_c,\n
&&a\ket{n}_c=\sqrt{n}\ket{n-1}_c,
\end{eqnarray}
the matrix elements of the Hamiltonian (\ref{TCB}) expressed in the invariant subspace (\ref{Invariant-subspace}) are given by
\begin{eqnarray}
\bm H_{jk}&\!=\!&\ex{j\!-\!1|_b\ex{n_0\!-\!j\!+\!1|_cH|k\!-\!1}_b|n_0\!-\!k\!+\!1}_c\n
&\!=\!&g\ex{j\!-\!1|_b\ex{n_0\!-\!j\!+\!1|_c(S^+a+{\rm H.C.})|k\!-\!1}_b|n_0\!-\!k\!+\!1}_c\n
&\!=\!&g\ex{j\!-\!1|_bS^+|k\!-\!1}_b\ex{n_0\!-\!j\!+\!1|_ca|n_0\!-\!k\!+\!1}_c\n
& &+g\ex{j\!-\!1|_bS^-|k\!-\!1}_b\ex{n_0\!-\!j\!+\!1|_ca^\dag|n_0\!-\!k\!+\!1}_c\n
&\!=\!&u_{j-1} \delta_{j-1,k}+u_{j}\delta_{j+1,k},
\end{eqnarray}
where $u_j\!=\!g\sqrt{j(N_b-k+1)(n_0-j+1)}$ and $j,k\!=\!1,2,\ldots,\min\{N_b,n_0\}$.

\section*{Appendix B: Generators of su(2) algebra}
The spin-$d$ irreducible representation of the generators of su(2) algebra is given by
\begin{eqnarray}
&&\bm S^x=\frac{1}{2}\begin{pmatrix}
0&f_1&\\
f_1&0&f_2&\\
&\ddots&\ddots&\ddots&\\
&&f_{d-1}&0&f_{d}\\
&&&f_{d}&0
\end{pmatrix},\n
&&\bm S^y=\frac{1}{2\rm i}\begin{pmatrix}
0&-f_1&\\
f_1&0&-f_2&\\
&\ddots&\ddots&\ddots&\\
&&f_{d-1}&0&-f_{d}\\
&&&f_{d}&0
\end{pmatrix},\n
&&\bm S^z=\begin{pmatrix}
-d&0&\\
0&-d\!+\!1&0&\\
&\ddots&\ddots&\ddots&\\
&&0&d\!-\!1&0\\
&&&0&d
\end{pmatrix}.
\end{eqnarray}
where $f_j\equiv\sqrt{j(2d-j+1)}$.
If we take $d\!=\!N_b/2$, then $\bm S^\alpha$ ($\alpha\!=\!x,y,z$) correspond to the operators $\bm J^\alpha$ [see Eqs. \,(\ref{H-limit-1}),\, (\ref{Energy-2}), and (\ref{BCH})].
If we take $d\!=\!n_0/2$, then $\bm S^\alpha$ correspond to the operators $\bm {\tilde J}^\alpha$ [see Eqs. \,(\ref{H-limit-2}) and (\ref{Energy-3})].

\bibliographystyle{apsrev4-1}
\bibliography{Quantum-Battery}

\end{document}